\documentclass[floats,floatfix,showpacs,amssymb,prd,superscriptaddress,twocolumn,aps,nofootinbib]{revtex4-1}
\usepackage{graphicx, epsfig, amssymb} 
\usepackage{amsmath, amsfonts}
\usepackage{bm} 

\usepackage[linktocpage]{hyperref}
\usepackage[caption=false]{subfig}
\usepackage[usenames]{color}

\newcommand{\dif}{{\rm d}}

\newcommand{\tu}{\tilde{u}}

\newcommand{\Lap}{\bigtriangleup}

\def\beq{\begin{equation}}
\def\eeq{\end{equation}}
\def\beqa{\begin{eqnarray}}
\def\eeqa{\end{eqnarray}}

\begin{document}
\title{\large On the nonlinear instability of confined geometries}
\author{Hirotada Okawa}
\affiliation{CENTRA, Departamento de F\'{\i}sica, Instituto Superior T\'ecnico, Universidade de Lisboa,
Avenida Rovisco Pais 1, 1049 Lisboa, Portugal}
\affiliation{Yukawa Institute for Theoretical Physics, Kyoto University, Kyoto 606-8502, Japan}
\affiliation{Advanced Research Institute for Science and Engineering, Waseda University, Tokyo 169-8555, Japan}

\author{Vitor Cardoso} 
\affiliation{CENTRA, Departamento de F\'{\i}sica, Instituto Superior T\'ecnico, Universidade de Lisboa,
Avenida Rovisco Pais 1, 1049 Lisboa, Portugal}
\affiliation{Perimeter Institute for Theoretical Physics Waterloo, Ontario N2J 2W9, Canada}

\author{Paolo Pani} 
\affiliation{CENTRA, Departamento de F\'{\i}sica, Instituto Superior T\'ecnico, Universidade de Lisboa,
Avenida Rovisco Pais 1, 1049 Lisboa, Portugal}
\affiliation{Dipartimento di Fisica, ``Sapienza'' Universit\`a di Roma, P.A. Moro 5, 00185, Roma, Italy}

\date{\today} 
\begin{abstract} 
The discovery of a ``weakly-turbulent'' instability of anti-de Sitter spacetime supports the idea that
confined fluctuations eventually collapse to black holes and suggests that 
similar phenomena might be possible in asymptotically-flat spacetime, for example in the context of 
spherically symmetric oscillations of stars or nonradial pulsations of ultracompact objects.
Here we present a detailed study of the evolution of the Einstein-Klein-Gordon system in a cavity, with different types of deformations of the spectrum, including a mass term for the scalar and Neumann conditions at the boundary. We provide numerical evidence that gravitational collapse always occurs, at least for amplitudes that are three orders of magnitude smaller than Choptuik's critical value and corresponding to more than $10^5$ reflections before collapse. The collapse time scales as the inverse square of the initial amplitude in the small-amplitude limit. 
In addition, we find that fields with nonresonant spectrum collapse \emph{earlier} than in the fully-resonant case, a result that is at odds with the current understanding of the process. Energy is transferred through a direct cascade to high frequencies when the spectrum is resonant, but we observe both direct- and inverse-cascade effects for nonresonant spectra.
Our results indicate that a fully-resonant spectrum might \emph{not} be a crucial ingredient of the conjectured turbulent instability and that other mechanisms might be relevant. We discuss how a definitive answer to this problem is essentially impossible within the present framework.
\end{abstract}

\maketitle

\section{Introduction}
Gravity has a complex and very rich nonlinear regime, as illustrated in Choptuik's classical 
study~\cite{Choptuik:1992jv}. Choptuik conducted a series of numerical experiments
to understand how black holes (BHs) form from an initial pulse of a massless, minimally coupled scalar field.
Large enough concentrations of matter collapse promptly, whereas
sufficiently ``small'' fluctuations eventually disperse to infinity in agreement with 
the nonlinear stability of Minkowski spacetime~\cite{1993gnsm.book.....C}.
For initial data described by an arbitrary parameter $p$, 
close to the threshold of BH formation the BH mass is described by~\cite{Choptuik:1992jv}
\beq
M\propto (p-p_*)^{\gamma}\,,\label{eq:mass_scaling}
\eeq
where $\gamma$ is a universal constant, independent on which parameter $p$ is used to describe the initial data, and $p_*$ is a critical value of $p$.

The nonlinear stability of Minkowski spacetime is ultimately due to the ability of small fluctuations to disperse;
what happens if such dissipation is instead negligible? For instance, in the assumption of vanishing viscosity, small spherical fluctuations of a star cannot be dissipated by gravitational waves due to Birkhoff's theorem, and are confined within the object.
Likewise, any fluctuation of a sufficiently compact object cannot be damped by gravitational waves and is trapped between the center of the object
and its unstable light ring~\cite{Keir:2014oka,Cardoso:2014sna}. A pertinent question is then what happens when Choptuik's experiment is repeated in \emph{confined} geometries?

A possible answer to this question was given by Bizon and Rostworowski, who investigated the collapse of scalar fields in asymptotically anti-de Sitter (AdS) spacetime~\cite{Bizon:2011gg}. Their study revealed that small enough fluctuations --~which would disperse to infinity in asymptotically-flat spacetime~--
bounce back and get (``weakly turbulent'') blueshifted upon gravitational interaction. In a detailed analysis, they argued that except for special sets of initial data, the evolution of a massless, spherically symmetric scalar field always results in collapse to a BH.
It has subsequently been suggested --~with strong numerical and analytic evidence to support it~-- that not all initial conditions lead to collapse~\cite{Buchel:2013uba,Balasubramanian:2014cja} (but see also~\cite{Bizon:2014bya}) but generic families of initial data remain that do collapse for arbitrarily small initial amplitudes\footnote{In this context by ``amplitudes''  we mean some parameter $p$ describing the initial data. In this work, we always refer to the amplitude $A$ of the initial wavepacket as defined in \eqref{eq:initPi_n} below.}.

\subsection{Open questions \& executive summary}
The results in AdS geometries raise the exciting -- if troublesome -- possibility that analogous effects might play a role in our universe. 
Indeed, stripping the process to its bare bones, Maliborski~\cite{Maliborski:2012gx} showed that the cosmological constant seems not be crucial: the evolution of a scalar field in a cavity leads to similar results as those found in Ref.~\cite{Bizon:2011gg}, i.e, some families of initial data collapse at arbitrarily small amplitudes. Nevertheless many --~and perhaps the most urgent~-- questions remain unanswered.
A fundamental limitation on our knowledge is due to the numerical nature of these studies, making it hard to decide on the trend of the process at very small amplitudes: lack of collapse for some finite initial amplitude doesn't necessarily mean that there is a threshold amplitude (because numerical simulations run for a finite time); likewise, collapse at small but finite amplitudes doesn't mean that the spacetime is necessarily nonlinearly unstable at {\it arbitrarily small} amplitudes. In this context some of the open questions are related to:

\noindent{{\em I. The (dynamical) approach to collapse.}} This question has been surfaced by estimating
how higher frequencies are excited, but no really solid picture of horizon formation has been attempted\footnote{For example,
which frequency distribution at the threshold of BH formation is guaranteed to give rise to a horizon?}. This work provides no
further insight on this issue.

\noindent{{\em II. The relevance of a fully-resonant spectrum.}} By using a perturbative expansion of the field equations, 
the original study~\cite{Bizon:2011gg} suggested that a fully-resonant --~or \emph{commensurable}~--
spectrum is a necessary condition for collapse at {\it arbitrarily small amplitudes}. We remind that commensurability is a condition on the proper eigenfrequencies of the background spacetime, i.e. there exist combinations of $i,j,k\in \mathbb{Z}$ such that the corresponding modes satisfy 
\beq
\omega_k=\omega_j-\omega_i\,,\quad k>j>i\,. 
\eeq
In a nutshell, the relation above implies the existence of secular terms that drive arbitrarily-small initial pulses away from the perturbative regime~\cite{Bizon:2011gg}.

Dias {\it et al.}~\cite{Dias:2012tq} have analyzed the full perturbative expansion and concluded that a resonant spectrum is indeed a necessary condition for collapse to occur at arbitrarily small amplitudes.  
Going beyond the perturbative level, some evidence exists that a nonresonant spectrum does halt collapse~\cite{Buchel:2013uba,Maliborski:2014rma}. Nevertheless, we believe that the existing numerical evidence is circumstantial and wish to present a thorough analysis of this problem. For instance, in Ref.~\cite{Maliborski:2014rma} only two points in parameter space are discussed. As we remarked previously, numerical simulations run for a finite amount of time and ``no collapse'' may simply mean that more
run time is necessary. In other words, we believe that \emph{trends} of collapse time versus wavepacket amplitude are necessary in order to make reliable statements. Such study is made in Ref.~\cite{Buchel:2013uba} (see e.g. Fig.~12), but again only for a limited region in parameter space. Finally, it is possible that such noncollapsing configurations are special, for example they might correspond to some ``island of stability''~\cite{Maliborski:2013jca,Buchel:2013uba}, which might not exclude the possibility that some other family of initial data collapses for arbitrarily small amplitudes.

Generically, compact geometries that can trap matter
do not give rise to a fully-resonant spectrum. Thus, understanding whether commensurability is necessary for weak gravitational turbulence
is mandatory. 
Here we show evidence that the fully nonlinear dynamics is much richer than what commensurability of the linearized perturbations would suggest.
In particular, by studying the evolution of a massive scalar in a cavity (which mimics in a simplistic way spherically symmetric perturbations of a star), we find evidence that collapse occurs at very small amplitudes with the collapse time scaling as the inverse of the square of the initial perturbation. We do not find evidence for a threshold amplitude, in contrast with the perturbative analysis of Ref.~\cite{Dias:2012tq}. We thus leave open the question of how small the initial amplitude needs to be for collapse to be prevented, or whether novel nonlinear mechanisms that cannot be captured by a perturbative analysis are responsible for collapse in nonresonant configurations (cf. Sec.~\ref{sec:relation} below for a detailed discussion).

\noindent{{\em III. Why nonresonant configurations collapse earlier.}} Interestingly --~although not explicitly discussed in previous work~-- numerical simulations
indicate that even for initial data that do not collapse at sufficiently small amplitude, the time for collapse at intermediate amplitudes
is of the same order of magnitude (or shorter) than those with fully-resonant spectrum (see Fig.~12 of Ref.~\cite{Buchel:2013uba}). Clearly, if collapse were driven by resonant excitations of higher frequencies,
the trend should be the opposite. Similar results can be inferred from the simulations reported in Ref.~\cite{Maliborski:2012gx}, where the time scales for resonant and nonresonant configurations are at least of the same order of magnitude throughout the amplitudes shown. Our results agree with and extend these findings, giving another indication for the existence of some other mechanism(s) driving collapse.

\noindent{{\em IV. Weak turbulence versus inverse cascade.}} In parallel with the perturbative analysis~\cite{Bizon:2011gg,Dias:2011ss,Dias:2012tq}, recent simulations have identified an inverse-cascade effect as the responsible for halting the collapse and leading to dynamical, nondispersive solutions~\cite{Balasubramanian:2014cja} (but see~\cite{Bizon:2014bya}). Large families of initial data exist whose late-time evolution is characterized by energy transfer to lower frequencies, thus being competitive to weak turbulence\footnote{An attempt to explain such behavior in terms of the existence of a stochasticity threshold has been recently proposed~\cite{Basu:2014sia}.}. Interestingly, in the case of a nonresonant configuration we also observe inverse-cascade effects, but the latter do not seem to prevent BH formation in all cases under consideration.

\noindent{{\em V. How small is the threshold amplitude.}} An equally relevant question concerns the threshold amplitude at which weakly-turbulent
collapse is possibly halted. In fact, whether or not collapse happens at arbitrarily small amplitudes is for some purposes an academic question,
as most physically interesting systems would be described by finite-sized perturbations. It is now established that weak-turbulence
causes collapse at amplitudes significantly lower than the Choptuik critical value. How much lower?
Our results show that in a confined geometry the critical amplitude for BH formation (if it exists) is at least three orders of magnitude smaller than Choptuik's critical amplitude for prompt collapse.

\noindent{{\em VI. Different matter, realistic configurations.}} Finally, the original simulations were performed solely for minimally coupled massless scalars in spherical symmetry~\cite{Bizon:2011gg,Maliborski:2012gx} (a notable exception is the analysis of Ref.~\cite{Dias:2011ss} dealing with nonspherical gravitational perturbations of AdS). If the process is to have any bearing at all on realistic configurations it is necessary to understand how these results generalize to other forms of matter. It is also not clear how nonlinearities will drive the system
in generic asymmetric configurations or in the presence of dissipation. We partially address this issue by considering massive scalars in our study, although retaining the assumption of spherical symmetry. (The role of dissipation in a partially-confining geometry has been investigated in Ref.~\cite{Okawa:2013jba}).

\section{Setup}
Our setup is similar to, but more general than, the one studied by other authors~\cite{Maliborski:2012gx}. We consider the Einstein-Klein-Gordon system for a massive scalar field $\Phi$, evolving in a spherically symmetric ``cavity''. The field is confined by an infinitely thin shell, the properties of which are determined a posteriori (see below), but which satisfy the dominant (and therefore the weak and the null) energy conditions. Outside the shell, the spacetime is asymptotically flat and vacuum. Because the system is spherically symmetric and the scalar is confined (in a way to be made precise below), the 
spacetime outside the shell belongs to the Schwarzschild family.

With these preliminaries, we consider the Einstein-Klein-Gordon action (henceforth we use $G=c=\hbar=1$ units)
\begin{align}
S = & \int d^4x \sqrt{-g} 
      \left( \frac{R}{16\pi} 
            -\frac{1}{2}\nabla^{\mu}\Phi\nabla_{\mu}\Phi -\frac{1}{2}\mu^2\Phi^2 \right)\,.\label{eq:action}
\end{align}
We focus on spherically symmetric spacetimes and our setup is the same as the one presented in Ref.~\cite{Okawa:2013jba} and briefly summarized here. In the Arnowitt-Deser-Misner decomposition, the geometry is described by
\begin{subequations}
 \label{eq:Ansatz_SS}
  \begin{align}
   \label{eq:Ansatz_SS_met}
   \dif s^2 =& -\alpha^2\dif t^2 +\psi^4\eta_{ij}\dif x^i\dif x^j\,,\\
   \label{eq:Ansatz_SS_ext}
   K_{ij} =& \frac{1}{3}\psi^4\eta_{ij}K\,,
  \end{align}
\end{subequations}
where $\eta_{ij}$ is the Minkowski 3-metric in spherical coordinates and $K_{ij}$ is the extrinsic curvature of the conformally flat metric $\gamma_{ij}=\psi^4\eta_{ij}$.

Variation of the action~\eqref{eq:action} gives the evolution equations
\begin{align}
 \dot\psi =& -\frac{1}{6}\alpha\psi K\,,\nonumber\\
 \dot K =& -\frac{1}{\psi^{4}}\Lap\hspace{-2pt}\alpha
 -\frac{2}{\psi^{5}}\psi'\alpha'
 +\frac{1}{3}\alpha K^2
 +4\pi\alpha\left(2\Pi^2 -\mu^2\Phi^2\right)\,,\nonumber\\
 \dot\Phi =& -\alpha\Pi\,,\nonumber\\
 \dot\Pi =& \alpha\Pi K -\frac{\alpha}{\psi^{4}}\Lap\hspace{-2pt}\Phi
 -\frac{1}{\psi^{4}}\alpha'\Phi'
 -\frac{2\alpha}{\psi^{5}}\psi'\Phi'
 \label{eq:evol_SS}
 +\alpha\mu^2\Phi\,,
\end{align}
where $\Pi$ is the conjugate momentum of $\Phi$, a dot and a prime
denote derivative with respect to $t$ and $r$,
respectively, and $\Lap$ is the flat Laplacian operator.
In addition, the Hamiltonian and momentum constraints read
\begin{align}
  \psi^{-5}\Lap\hspace{-2pt}\psi 
 -\frac{K^2}{12}
 +\pi \left[\Pi^2 +\psi^{-4}\Phi'^2+\mu^2\Phi^2\right] &=&\; 0\,,\\
 \label{eq:Constr_SS}
 \frac{2}{3}K' +8\pi\Pi\Phi' &=&\; 0\,,
\end{align}
where the kinetic term $\Lap\psi$
can be computed by the so-called CARTOON method~\cite{Alcubierre:1999ab}
described in the Appendix~\ref{app:cartoon}. The evolution system is
closed by prescribing a gauge condition for the lapse function and we
use the so-called ``1+log'' slicing condition, $\dot{\alpha}=-2\alpha K$.
The latter equation, together with Eqs.~\eqref{eq:evol_SS} can be solved by implementing a free evolution scheme. We note that such scheme might drive the evolution
away from the true solution of the initial value problem, even when standard convergence criteria are met. This behavior can be prevented by introducing a fictitious damping term. To exclude this issue, we have instead checked \emph{a posteriori} that our solutions do solve the constraints~\eqref{eq:Constr_SS} and the evolution equations at all times, especially for long evolutions. 

\subsection{Boundary conditions}
Regularity at the origin $r=0$ is guaranteed by imposing Neumann conditions for all variables $\alpha,\psi,K,\Phi,\Pi$. 
On the other hand, imposing physically-motivated boundary conditions at the surface of the cavity, $r=R$, is a nontrivial task in a fully nonlinear evolution. Generically, the boundary of the cavity contributes to the dynamics with its own stress-energy tensor, it can oscillate and resonate with the perturbations, etc. 
Here, we will deal with the boundary by immersing it in a vacuum, asymptotically-flat spacetime, and take the boundary to be composed of
an infinitely thin shell of some material which obeys some relevant energy conditions.
\subsubsection{Dirichlet and Neumann boundary conditions}
First of all, it is useful to define a ``mass function'' $m(r,t)$ for the spacetime through $\psi(r,t)=1+m(r,t)/2r$. Although $m(r,t)$ has no particular meaning, if one embeds the cavity in an asymptotically-flat spacetime and confines the scalar within the cavity, Birkoff's theorem guarantees that the exterior spacetime has to be described by the Schwarzschild metric with $m(r,t)=M_S$ for $r>R$. As we discuss in a moment, the properties of the cavity surface are related to the discontinuity of the metric functions. Therefore, it is natural to impose that $m(r,t)$ be constant in time as $r\to R$ in order to guarantee that the intrinsic properties of the shell remain also constant. 

From the evolution equations we get
\begin{eqnarray}
 \dot m &=& 2r\dot\psi = -\frac{1}{3}r\alpha\psi K\,,
\end{eqnarray}
and therefore (discarding singular situations where $\alpha$ or $\psi$ are zero at $r=R$) the mass $M\equiv m(R,t)$ is constant if and only if $K(R,t)=0$. In the following we impose the latter condition to ensure that $m(R,t)$ is a conserved quantity and we shall check \emph{a posteriori} that this condition is enough to guarantee that the shell's properties do not change in time.

In order to characterize the behavior near the boundary, we expand each variable (collectively denoted as $X$) near $r=R$ as
\begin{equation}
 X(r,t)=\sum_{i=0}^N X_i(t)\left(1-\frac{r}{R}\right)^i\,.
\end{equation}
Finally, we impose two sets of boundary conditions:
\begin{eqnarray}
{\rm Dirichlet:}\quad \Phi_0(t) &=&\Pi_0(t) =K_0(t)=0\,,\label{DirichletBC}\\
{\rm Neumann:}\quad \Phi_1(t) &=&\Pi_1(t) =K_0(t)=0. \label{NeumannBC}
\end{eqnarray}
both supplemented by the gauge condition $\alpha_0(t)=1$ near the boundary.

By virtue of the field equations, both sets of boundary conditions imply that the metric coefficients are constant at the boundary. Indeed, by imposing Dirichlet boundary conditions~\eqref{DirichletBC} and by using the field equations, we obtain
\beq
K_1(t)=\Pi_1(t)=0\,, \alpha_1(t)=\alpha_1\,,\psi_0(t)=\psi_0\,,\psi_1(t)=\psi_1\,,\nonumber
\eeq
where $\alpha_1$, $\psi_0\equiv 1+M/(2R)$ and $\psi_1$ are constants fixed by the initial data. All the remaining functions $X_i$ can be written in terms of these constants by solving the field equations iteratively near the boundary.

Likewise, Neumann boundary conditions~\eqref{NeumannBC} imply
\beqa
K_1(t)&=&\Pi_0(t)=0\,,\nonumber \\
\alpha_1(t)&=&\alpha_1\,,\psi_0(t)=\psi_0\,,\psi_1(t)=\psi_1\,,\Phi_0(t)=\Phi_0\,,\nonumber
\eeqa
where $\Phi_0$ is also constant by virtue of the field equations.

\subsubsection{The spectrum of perturbations}
To first order, the scalar field satisfies a massive Klein-Gordon wave equation propagating on a Minkowski background, $\square_0 \Phi_1=\mu^2\Phi_1$, whose solution can be decomposed in normal modes:
\begin{equation}
 \Phi_1(r,t)=\sum_i a_i \cos(\omega_i t+b_i)e_i(r) \label{phi10}
\end{equation}
where $a_i$ and $b_i$ are constants and the orthonormal eigenfunctions that are regular at the center read
\begin{equation}
e_i(r)=B_i\sqrt{\frac{2}{R}}\frac{\sin(k_i r)}{r}\,.
\end{equation}
Here we have defined the norm $\langle e_i,e_j\rangle\equiv \int_0^R dr r^2 e_j e_i$, and $B_i=1,\,\sin^{-1}(k_iR)$ for Dirichlet and Neumann conditions, respectively.
The dispersion relation for a massive wave is simply $\omega_i^2=k_i^2+\mu^2$. 
For a massive scalar field enclosed in a spherical cavity, the spectrum reads
\begin{eqnarray}
 \omega_j&=&\sqrt{\mu^2+({j\pi}/{R})^2}\,, \quad \text{Dirichlet} \label{modesD0}\\
 \omega_j&=&\sqrt{\mu^2+\hat{k}_j^2}\,,    \quad \hspace{0.85cm}\text{Neumann}  \label{modesN0} 
\end{eqnarray}
for Dirichlet and Neumann boundary conditions at $r=R$, respectively, and where $\hat{k}_j$ ($j=0,1,2,...$) are the roots of $\tan(kR)=kR$. 
Therefore, the modes of the linearized problem are commensurable only in the massless case when Dirichlet conditions are imposed at the boundary.

\subsubsection{Junction conditions}

With the boundary conditions at hand, we can use Israel's junction conditions~\cite{Israel:1966rt} across $r=R$ to compute the properties of the cavity surface. Because the metric coefficients are constant at $r=R$, the line element~\eqref{eq:Ansatz_SS_met} can be written in Schwarzschild-like form near $r=R$ as
\begin{equation}
 ds^2=-\alpha^2 \left(1-\frac{2M_S}{y_R}\right)dT^2+\frac{dy^2}{F^2}+y^2(d\vartheta^2+\sin^2\vartheta d\varphi^2)\,,
\end{equation}
where $y=r\psi^2$, $y_R=y(R)$, $t=T(1-2M_S/y_R)$, $F=1+2r\psi'/\psi$ and $M_S$ is the Arnowitt-Deser-Misner mass of the asymptotically-flat spacetime in which the cavity is embedded. This solution has to be matched with the standard Schwarzschild solution with mass $M_S$ across the cavity surface ${\cal S}$ at $r=R$. Regularity of the intrinsic three-metric requires $[[g_{AB}]]=0$, where $[[X]]=X_+-X_-$ is the jump across the surface of a given quantity $X$ and the indices $A,B$ run over the coordinates $T$, $\vartheta$ and $\varphi$ that parametrize curves tangential to ${\cal S}$. The condition $[[g_{AB}]]=0$ is automatically satisfied in the coordinates $(T,y,\vartheta,\varphi)$. The remaining junction conditions stem from Einstein's equations and read
\begin{equation}
 [[{\cal K}_{AB}]]=8\pi(S_{AB}-\gamma_{AB}S/2)\,,
\end{equation}
where ${\cal K}_{AB}$ is the extrinsic curvature of ${\cal S}$, $\gamma_{AB}$ is the induced three-metric on ${\cal S}$ and  $S_{AB}=(\Sigma-\Theta)u_A u_B-\Theta \gamma_{AB}$ is the surface-energy tensor, with $\Sigma$ and $\Theta$ being the surface energy and the surface tension of the cavity, respectively, and $u_A$ being the fluid $3$-velocity vector. In our spherically symmetric case $\Sigma$ and $\Theta$ are related to the jump of the extrinsic curvature through 
\begin{eqnarray}
\Sigma=-\frac{[[{\cal K}^\vartheta_\vartheta]]}{4\pi}\,, \qquad \Theta=-\frac{\Sigma}{2}+\frac{[[{\cal K}^t_t]]}{8\pi}\,.
\end{eqnarray}
where $[[{\cal K}_{AB}]]=[[\sqrt{g^{yy}}g_{AB,y}/2]]$ (cf. e.g.~\cite{Visser}). It is now clear that having the metric and its derivatives constant at $r=R$ simply translates in the fact that $\Sigma$ and $\Theta$ are constant in time. In general, $\Sigma$ and $\Theta$ depend on the values of $\alpha_1$, $\psi_0\equiv 1+M/(2R)$, $M_S$ and $\psi_1$. Equivalently, once the initial data and the shell's surface energy $\Sigma$ are prescribed, the junction conditions allow to determine the total mass $M_S$ and the shell's equation of state, $\Theta(\Sigma)$.

In the following we shall consider initial data corresponding to $\alpha_1=0$, $\psi_0=1+{\cal O}(\epsilon^2)$, $\psi_1={\cal O}(\epsilon^2)$, where $\epsilon$ is related to the amplitude of the initial data. In this case we obtain
\begin{eqnarray}
\Sigma\sim \frac{1-\sqrt{1-\frac{2 M_S}{R}}}{4 \pi  R}\,,\qquad \Theta&\sim&-\frac{\Sigma}{2}\frac{R-M_S}{R-2M_S}\,,
\end{eqnarray}
in the small-$\epsilon$ limit, for both Dirichlet ($\Phi_0=0$) and Neumann boundary conditions. As it can be checked, the shell's surface satisfies the dominant energy condition for $R>3M_S$, and it violates the strong energy condition for any $R>2M_S$.

\begin{figure*}[ht]
 \begin{tabular}{cc}
  \psfig{file=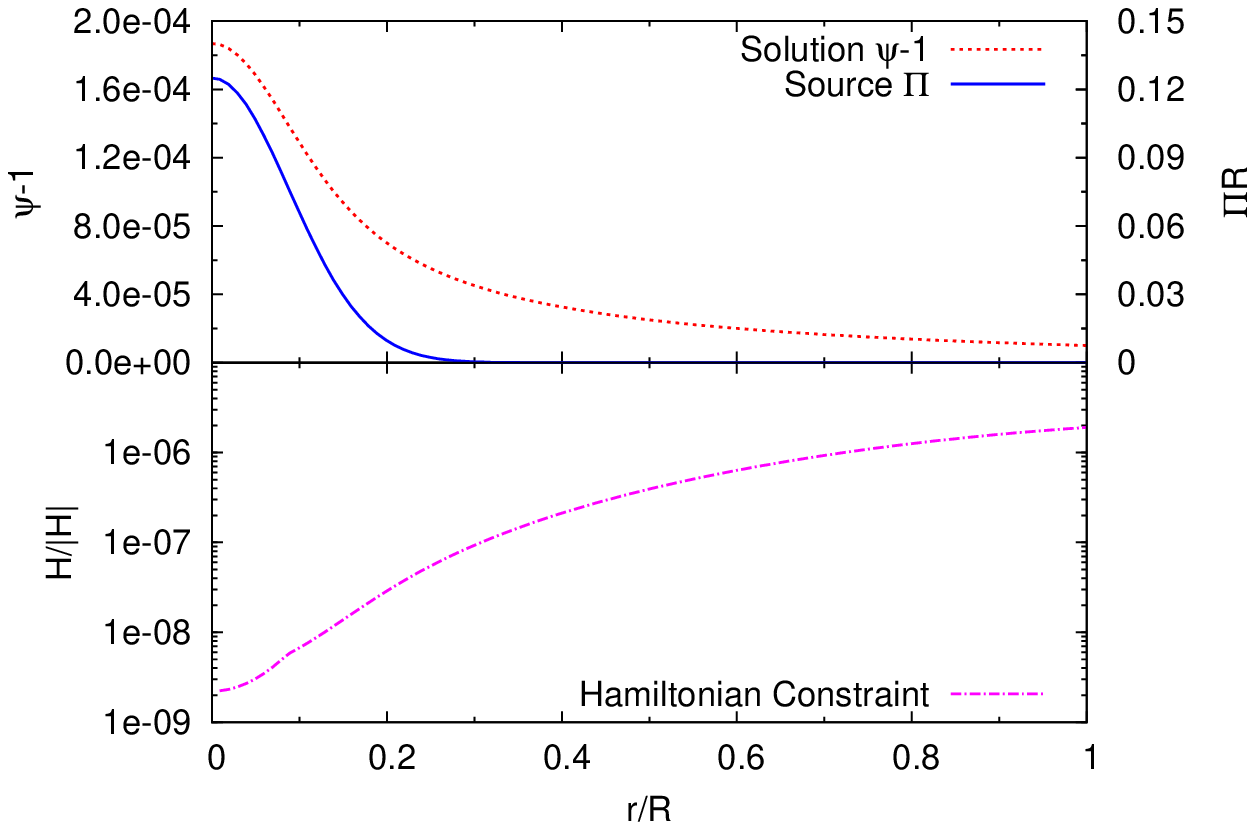,width=8.5cm}&
  \psfig{file=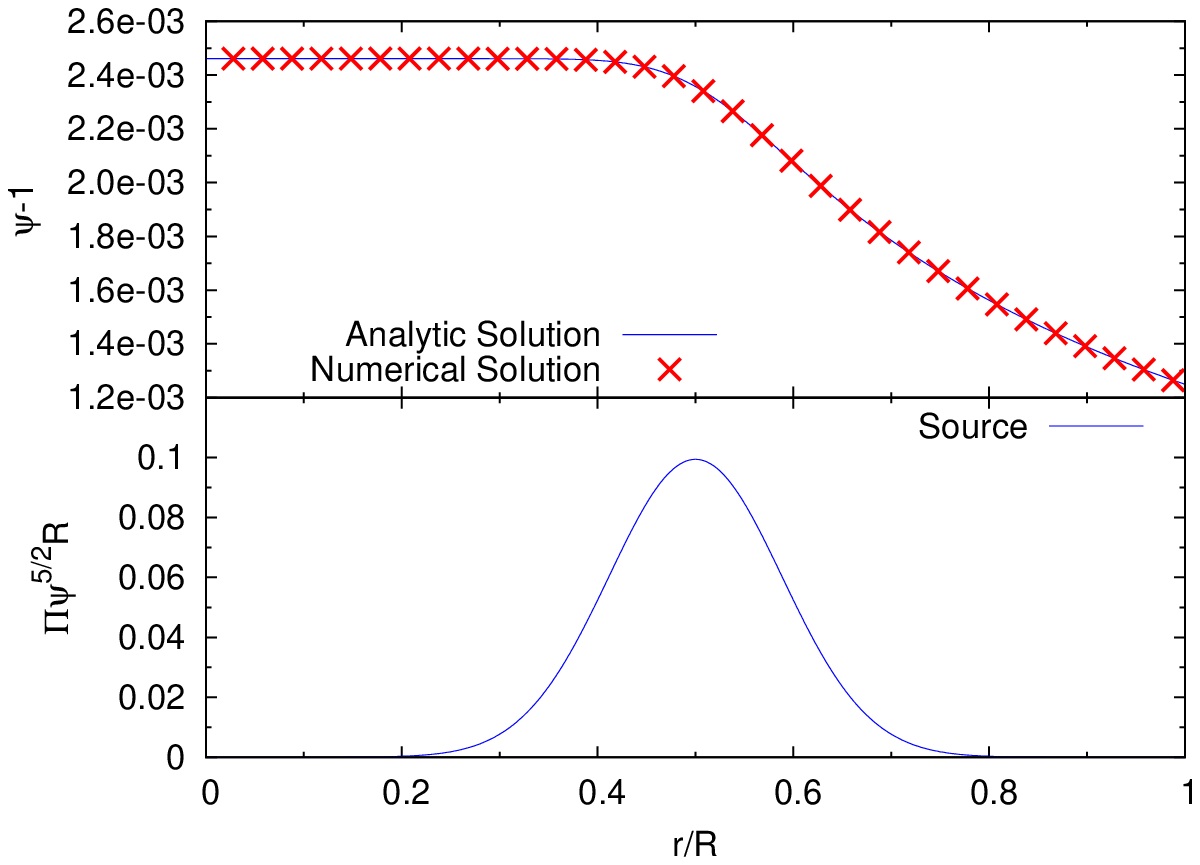,width=8.5cm}
 \end{tabular}
 \caption[]{
 Left panels: Example of numerical initial data. The initial profile is given by Eq.~\eqref{eq:initPi_n}
 with $AR=0.125$ and $w=0.125R$.
 The left top panel shows the source $\Pi$ and the conformal factor as computed numerically with resolution $dr=R/1000$. The left bottom panel shows the normalized error of the Hamiltonian constraint at $t=0$. Right panels: Comparison between analytic and numerical initial data. In the right top panel we compare the numerical and analytic solution
 for a source described in Eq.~\eqref{eq:initPi_a} with $AR=0.1$ and $w=0.125R$. The bottom right panel shows the source profile including the conformal factor.
 }
 \label{fig:initdata}
\end{figure*}
%
\subsection{Diagnostics}
We will make extensive use of two diagnostic tools. The first diagnostic is a proxy for BH formation via identification
of an apparent horizon, a condition for which is that~\cite{Thornburg:2006zb}
\beq
\partial_r(r^2\psi^4)=\frac{2}{3}r^2\psi^6K\,.
\eeq

The second diagnostic, identifying strong curvature regions, consists on evaluating the following Ricci and Kretschmann scalars at the origin, where a small BH and regions of large curvature are expected to form,
\beqa
g^{\mu\nu}R_{\mu\nu}&=&8\pi\left(-\Pi^2+2\mu^2\Phi^2\right)\,,\\
R_{\alpha\beta\gamma\delta}R^{\alpha\beta\gamma\delta}&=& \frac{2}{27}\left[K^4-24\pi K^2\left(\Pi^2+\mu^2\Phi^2\right)\right]+\frac{8(\alpha'')^2}{3\alpha^2\psi^8}\nonumber\\
&+&\frac{8}{3}\left[4\pi^2\left(11\Pi^4-2\mu^2\Pi^2\Phi^2+5\mu^4\Phi^4\right)\right].
\label{scalar_curvature}
\eeqa
In our formulation, $\Phi$ and $\Pi$ remain generically bounded and regular at the origin and as a consequence so does the Ricci scalar. However,
the first two terms in the Kretschmann, respectively proportional to $K^4$ and $K^2$, will be seen to cause an ultra-exponential growth
in the scalar curvature at the center\footnote{Using the metric ansatz of Ref.~\cite{Maliborski:2012gx}, the Kretschmann scalar at the origin reads $R_{\alpha\beta\gamma\delta}R^{\alpha\beta\gamma\delta}\propto (g^{\mu\nu}R_{\mu\nu})^2$. In this case the Ricci curvature (and therefore the Kretschmann scalar) grows ultra-exponentially at the origin when an apparent horizon is formed. Although the Ricci curvature is an invariant, this behavior is not necessarily in contrast with the fact that our Ricci scalar is bounded at $r=0$. We suspect that the difference is due to the different slicing and coordinate patches used in the two cases (note that in our coordinate $r=0$ does not necessarily correspond to the origin of the coordinates used in Ref.~\cite{Maliborski:2012gx} and that even for the Schwarzschild solution $g^{\mu\nu}R_{\mu\nu}=0$ at the origin whereas the Kretschmann is singular). Modulo this difference, the endstate of the evolution is consistent in the two cases, as discussed in the rest.}.

\subsection{Initial data}
We consider two sets of initial data that solve the constraints~\eqref{eq:Constr_SS}:
(i) numerical initial data similar to the ones of Ref.~\cite{Maliborski:2012gx}
and (ii) the same analytic initial data as those described in Ref.~\cite{Okawa:2013jba}.
In both cases, we assume $K=0=\Phi$ at $t=0$, so that
the momentum constraint in Eq.~\eqref{eq:Constr_SS} is trivially satisfied.
Then, for a given profile $\Pi(0,r)$, the conformal factor $\psi\equiv 1+u$
 is determined by the Hamiltonian constraint,
\begin{eqnarray}
 \Lap\hspace{-2pt}u &=& -\pi\Pi^2\psi^5\,, \label{eq:initialconformal}
\end{eqnarray}
which we solve with boundary conditions $\partial_r(ru)=0$ at $r=R$ and $\partial_ru=0$ at $r=0$. 
In this work we consider the following profiles for $\Pi$:
\begin{subequations}
 \begin{align}
  \label{eq:initPi_n}
  \Pi(0,r) =& A \exp\left\{-r^2/w^2\right\},\\
  \label{eq:initPi_a}
  \Pi(0,r) =& A \exp\left\{-(r-R/2)^2/w^2\right\}\psi^{-5/2},
 \end{align}
\end{subequations}
where $A$ and $w$ denote the amplitude, width and location of
the Gaussian wave packet. In the first case Eq.~\eqref{eq:initialconformal} is solved numerically
by discretization and implementing an iterative method,
\begin{eqnarray}
 u_{j} &=& \frac{1}{2}\left( u_{j+1} +u_{j-1} \right)
  +2\left( \tu_{j} -u_{j} \right) +\frac{1}{2}\pi\Pi^2_j(1+u_j)^5 dr^2,\nonumber
\end{eqnarray}
where $\tu_{j}$ is obtained by the CARTOON method in order to avoid large numerical errors near the center, see Appendix~\ref{app:cartoon}. 
We also use the Multi-Grid method to efficiently reduce the numerical
error during the iteration. 

On the other hand, by using Eq.~\eqref{eq:initPi_a}, a regular solution for the Hamiltonian constraint~\eqref{eq:initialconformal} can be found in closed form.
We refer to Refs.~\cite{Okawa:2013jba,Okawa:2014nda} for details on the analytic initial data.

In the left panel of Fig.~\ref{fig:initdata} we show an example of numerical solution
for the initial profile~\eqref{eq:initPi_n}, whereas in the right panel of Fig.~\ref{fig:initdata} we compare the analytical initial data for the initial profile~\eqref{eq:initPi_a}
used in Ref.~\cite{Okawa:2014nda} with the numerical initial data for the same profile.

\section{Results}
We performed various simulations for different values of the scalar-field
mass, different boundary conditions and various types of initial data; for each of these simulations we computed the time of apparent horizon formation for different initial pulse amplitudes, and we followed the growth of the scalar curvature \eqref{scalar_curvature} at the center. The results are discussed in the
following and summarized in Figs.~\ref{fig:coltime_Dirichlet}-\ref{fig:upward}. All quantities are normalized by the cavity size $R$.
The convergence properties of our results are detailed in Appendix \ref{app:convergence} and are consistent with
the adopted numerical procedure.

When the initial amplitude is larger than a critical value $A_*$, BH formation occurs promptly as in the previous mentioned Choptuik's study~\cite{Choptuik:1992jv} 
in asymptotically-flat space (which we have recently investigated in detail for massive fields~\cite{Okawa:2013jba}). 
Here, we are particularly interested in the behavior of the system for
amplitudes much smaller than Choptuik's critical value, for which it takes a large number of reflections
at the cavity boundary before horizon formation.

\subsection{Dirichlet boundary}
%
\begin{figure}[t]
 \psfig{file=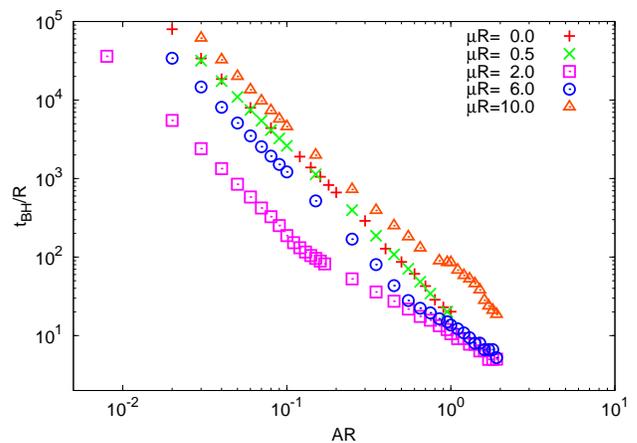,width=8.5cm}
 \caption[]{Collapse time as a function of initial amplitude $A$ with
 Dirichlet boundary conditions imposed at $r=R$, pulse width
 $w=0.125R$ and various scalar field masses. For
 this initial data~\eqref{eq:initPi_a}, Choptuik's critical amplitude for prompt collapse
 corresponds to $A R\sim 8$.
 }
 \label{fig:coltime_Dirichlet}
\end{figure}
As we already mentioned, Dirichlet boundary conditions yield a fully-resonant spectrum for massless perturbations
but not for generic massive fields. As a first exploration of the impact of a commensurable spectrum, we have therefore
evolved massive fields subjected to Dirichlet conditions~\eqref{DirichletBC} at $r=R$. Results for the collapse time $t_{\rm BH}$ 
are shown in Fig.~\ref{fig:coltime_Dirichlet} as a function of the initial amplitude for various masses $\mu$.

For massless fields, $\mu=0$, we confirm the results of Ref.~\cite{Maliborski:2012gx},
which found evidence for a weakly-turbulent instability for generic initial data,
similar to the AdS case~\cite{Bizon:2011gg}
and in agreement with the fact that the spectrum of linear perturbations is fully resonant.
In this case the collapse time $t_{\rm BH}\sim1/A^2$ at small amplitudes and for different initial data,
in agreement with previous studies~\cite{Bizon:2011gg,Maliborski:2012gx}.

As shown in Eq.~\eqref{modesD0}, commensurability of the mode spectrum in
the Dirichlet case is broken for any nonvanishing mass. Should a fully-resonant spectrum be a necessary condition for BH formation starting from arbitrarily small initial data, one would expect the appearance of a critical amplitude in the function $t_{\rm BH}(A)$. In other words, Fig.~\ref{fig:coltime_Dirichlet} should display a qualitatively different behavior for the massive case relative to the massless case, with 
$t_{\rm BH}\to\infty$ as the critical amplitude is approached (cf. e.g. Figs.~7 and 8 in Ref.~\cite{Buchel:2013uba}).
However, Fig.~\ref{fig:coltime_Dirichlet} does not indicate
any qualitatively different behavior for $0\leq\mu R\leq10$.
Choptuik's critical amplitude for the case shown in Fig.~\ref{fig:coltime_Dirichlet}
corresponds to $A_* R\sim 8$ and, therefore,
our simulations show that collapse still occurs for amplitudes
that are {\it three orders of magnitude} smaller than $A_*$.
This corresponds to roughly $10^5$ reflections from the cavity boundary
before the formation of an apparent horizon.
In the region under consideration our results are still compatible
with a power-law dependence, $t_{\rm BH}(A)\sim A^{-2}$ at small amplitudes and for any mass.

Even more interestingly, the collapse time for $\mu R\lesssim 1$ is always of the same order of magnitude or \emph{shorter} than the collapse time in the massless case for the same initial data (i.e., fixing the amplitude and width of the initial pulse). This result is counter-intuitive in terms of perturbation theory, because one would expect that the fully-resonant spectrum of the massless case is the most favorable situation to trigger nonlinear effects and lead to BH formation.

\begin{figure}[t]
 \psfig{file=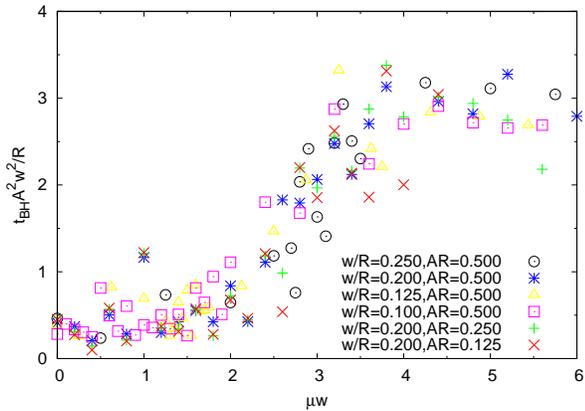,width=8.cm}
 \caption[]{Collapse time vs ratio of Compton wavelength to the width of initial pulse ($\mu w$) for Dirichlet boundary.
 Initial data are described in \eqref{eq:initPi_n} with $dr=R/2000$.
 }
 \label{fig:coefficient}
\end{figure}

When the Compton wavelength of the scalar field is considerably smaller than any other lengthscale in the problem 
($\mu R\,,\mu w\gg1$), the coefficient of $t_{\rm BH}(A)\sim A^{-2}$ is a nonmonotonic function of $\mu$ and, indeed, the collapse time for $\mu R=10$ is longer than in the massless case for the same amplitude, although it does not show deviations from the power-law behavior at small amplitudes.

A study of the collapse time for different mass parameter $\mu w$ is shown in Fig.~\ref{fig:coefficient} and shows three important features:
(i) the collapse time $t_{\rm BH}$ does scale like the inverse of the square amplitude for basically all parameters we studied, but (ii)
there is a transition in the coefficient of proportionality at the point where the Compton wavelength becomes of the order of the wavepacket width. Furthermore, (iii) the collapse time increases as the width $w$ decreases. This is exactly the opposite as for the case of prompt collapse~\cite{Choptuik:1992jv}. We shall come back to this point later.

\subsection{Neumann versus Dirichlet boundary}
%
\begin{figure}[t]
 \psfig{file=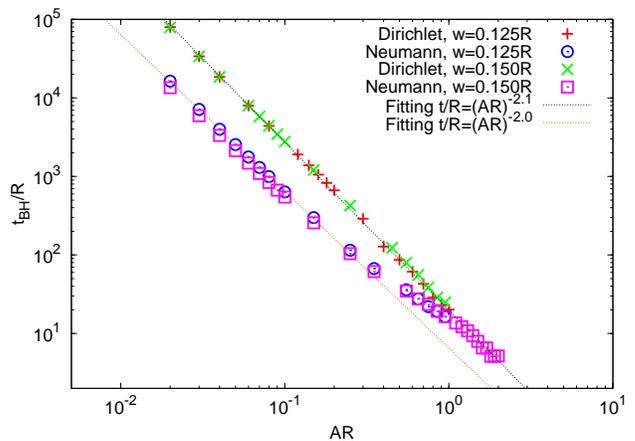,width=8.5cm}
 \caption[]{Collapse time as a function of initial amplitude with
 Neumann boundary conditions and compared with Dirichlet boundary
 conditions for a massless scalar field.
 Initial data are the same in both cases: we considered the initial pulse~\eqref{eq:initPi_a}
 with two choices of the width, namely $w=0.125R$ and $w=0.15R$.
 }
 \label{fig:coltime_DvsN}
\end{figure}
As another relevant example in which the mode spectrum is dispersive, we considered Neumann boundary
conditions~\eqref{NeumannBC}, which correspond to the nonresonant spectrum~\eqref{modesN0} at the linearized level. As shown in Fig.~\ref{fig:coltime_DvsN}, also in this case we observe an approximate power-law behavior of the collapse time as a function of the initial amplitude, and our simulations do not show deviations from this behavior in a range of amplitudes that covers almost three orders of magnitude.

It is also interesting to compare the Neumann and the Dirichlet cases. Again, the absence of a fully-resonant spectrum in the former would suggest that the collapse time should be longer than the corresponding collapse time obtained evolving the same initial data with Dirichlet boundary conditions. On the contrary, Fig.~\ref{fig:coltime_DvsN} shows that the time of BH formation in the Neumann case is always of the same order of magnitude or \emph{shorter} for a given initial amplitude, similarly to the massive case (with $\mu R\lesssim 1$) discussed above. Similar results for Neumann conditions
can be extracted from the simulations of Ref.~\cite{Maliborski:2012gx}.

\begin{figure}[ht]
 \psfig{file=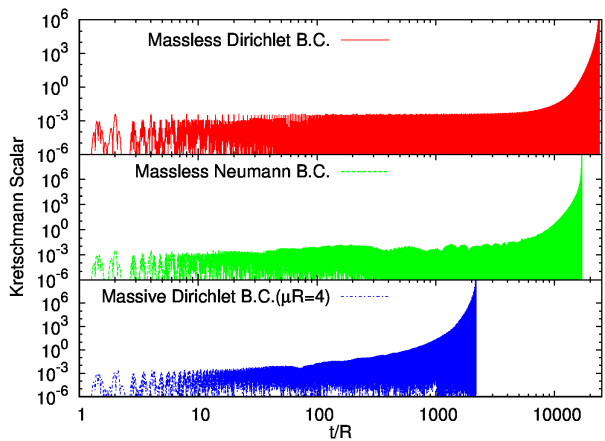,width=8.5cm}
 \caption[]{The evolution of the Kretschmann scalar for three
 representative cases with resolution $dr=R/3200$: a massless scalar field with Dirichlet boundary
 (top panel) and with Neumann boundary (middle panel) and a massive
 scalar field with $\mu R=4$ and Dirichlet boundary. In all cases
 initial data are prescribed through Eq.~\eqref{eq:initPi_n}, $A R=0.04$
 and $w=0.125R$. The curves are truncated at the time of
 formation of an apparent horizon. Note that the case corresponding to a
 fully-resonant spectrum is the one for which collapse takes longer. Our results show that the Kretschmann scalar at the origin scales with the fourth-power of the initial amplitude, $A^4$, at least for small amplitudes and well before horizon formation.
 \label{fig:curvature}
 }
\end{figure}

The formation of an apparent horizon in our simulations is associated with regions
of large curvature, indicating the formation of a small BH. This is clearly seen in Fig.~\ref{fig:curvature}, where
we show the Kretschmann scalar at the origin (c.f. Eq.~\eqref{scalar_curvature}) as a function of time in three cases:
(i) massless case with Dirichlet boundary, (ii) massless case with Neumann boundary
and (iii) massive case ($\mu R=4$) with Dirichlet boundary. The curves are truncated at the time of formation of an apparent horizon.
In all cases under consideration the evolution is characterized by an ultra-exponential growth of the Kretschmann scalar near BH formation.
This signals the formation of a large curvature region, consistent with the previous finding of formation of a small BH.
This is similar to the divergence of the Ricci curvature of the geometry shown in Refs.~\cite{Bizon:2011gg,Maliborski:2012gx}. 
These particular plots refer to the same initial wavepacket amplitude and width, and therefore strengthen the statement
that some mechanism other than mode resonance is at play: fields with Neumann conditions or with a mass term collapse earlier
and the ultra-exponential growth in the curvature also happens earlier than for (massless, Dirichlet) fields with a resonant spectrum.

\begin{figure*}[t]
\centering
\begin{minipage}[t]{.48\textwidth}
\psfig{file=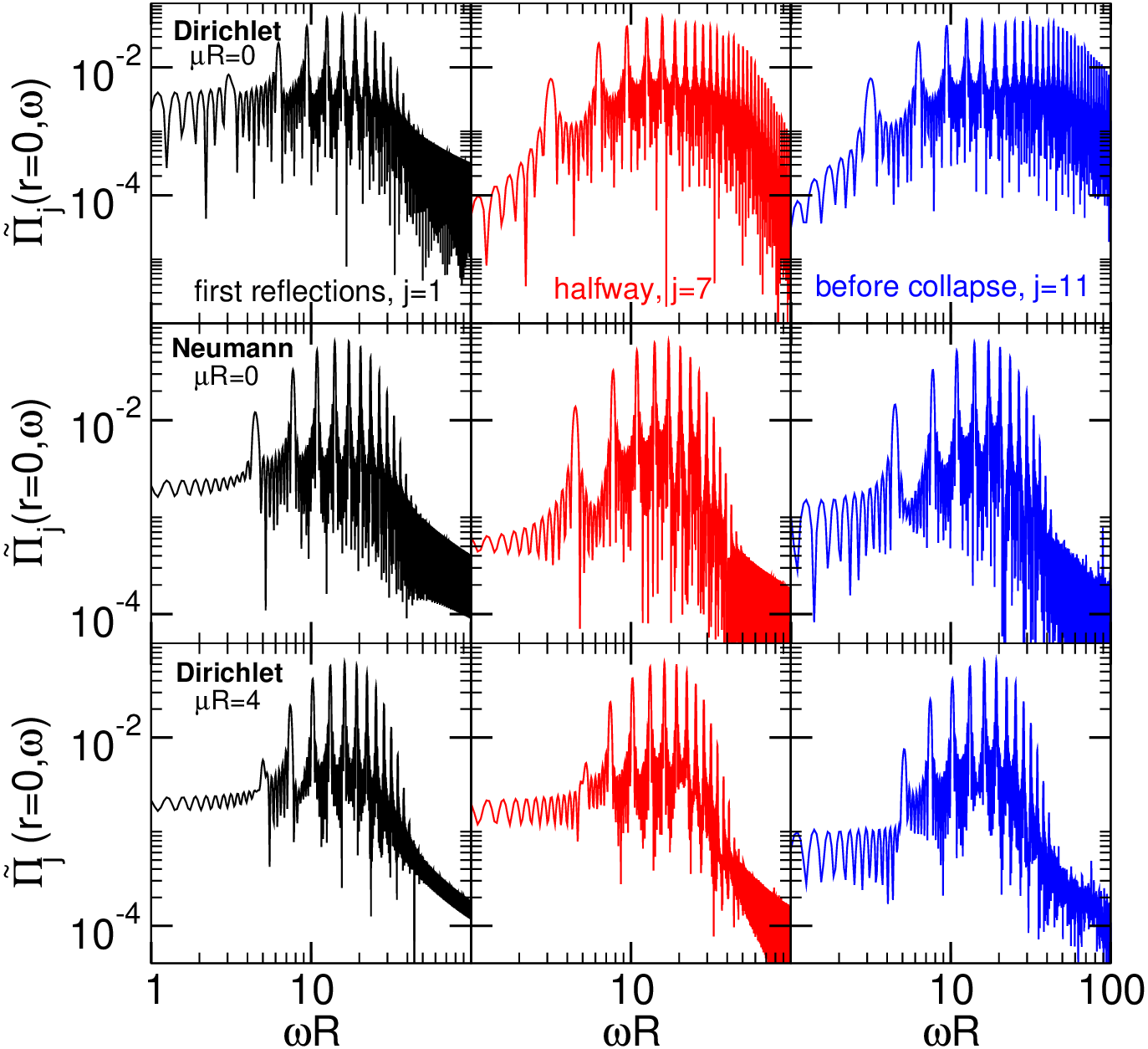,width=8.5cm}
 \caption[]{Fourier-transform $\tilde{\Pi}_j$ of the conjugate momentum $\Pi$ at the origin for different time intervals. We consider three intervals corresponding to the first 10 reflections ($j=1$, black curves, left panels), last 10 reflections before the collapse ($j=11$, blue curves, right panels), and for 10 reflections roughly centered at $t=t_{\rm BH}/2$ ($j=7$, red curves, middle panels). The top, central and bottom panels show the case of massless fields with Dirichlet boundary, massless field with Neumann boundary and massive field with Dirichlet boundary, respectively. 
 All cases refer to initial data~\eqref{eq:initPi_n} with $AR=0.04$ and $w=0.125R$.
}
\label{fig:FT}
\end{minipage}\hfill
\begin{minipage}[t]{.48\textwidth}
\psfig{file=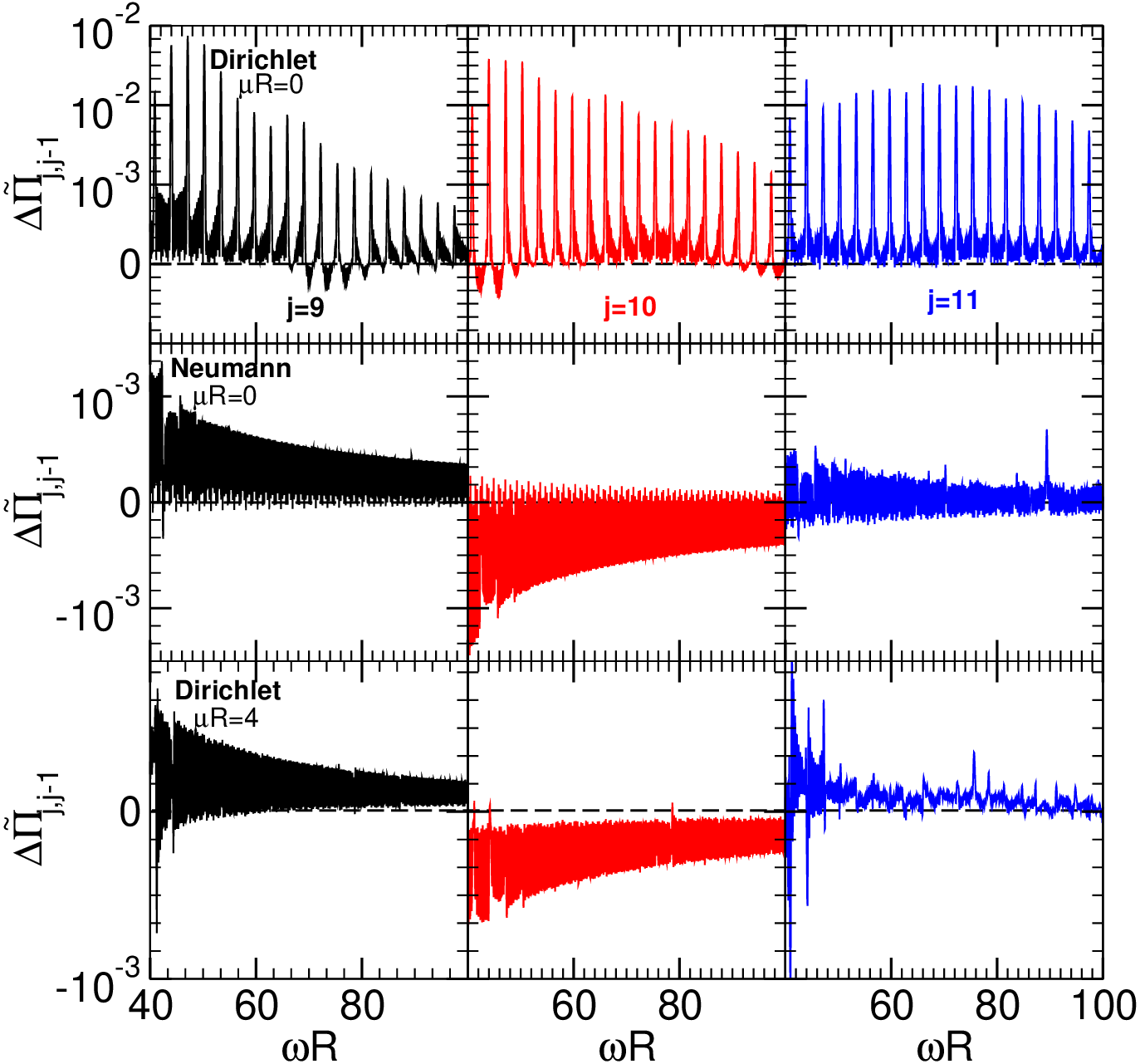,width=8.5cm}
\caption[]{Direct and inverse cascade effects. The difference $\Delta\tilde{\Pi}_{j,j-1}\equiv \tilde{\Pi}_j-\tilde{\Pi}_{j-1}$ at the center between two consecutive time intervals is shown in the high-frequency region $40\leq\omega R\leq100$. Left, middle and right panels correspond to different time intervals close to the formation of the apparent horizon (from left to right, $j=9,10,11$).  
The top, middle and bottom panels show the case of massless fields with Dirichlet boundary, massless field with Neumann boundary and massive field with Dirichlet boundary, respectively. In the massless Dirichlet case the difference is always positive and the eigenmodes are clearly excited, confirming a direct-cascade effect. In the Neumann and Dirichlet massive cases we observe both direct and inverse cascade.
All cases refer to initial data~\eqref{eq:initPi_n} with $AR=0.04$ and $w=0.125R$.}
\label{fig:FT_cascade}
\end{minipage}
\end{figure*}
\subsection{Direct- and inverse-cascade effects}

Since the pioneering work on the nonlinear instability of AdS~\cite{Bizon:2011gg}, BH formation in confined geometries is associated with a weakly-turbulent instability: during the evolution the energy is shifted to higher and higher frequencies thus probing smaller and smaller scales and eventually forming an apparent horizon~\cite{Bizon:2011gg}. To investigate this phenomenon, we have taken 11 samples $\Pi_j,\,j=1,...,11$ of the conjugate momentum at equispaced time intervals, with each sample containing 10 reflections (or in other words each of these 10 intervals lasts for $20R$); the results are not qualitatively dependent on the size of these samples, as long as they are much smaller than the total simulations time. We then Fourier-transformed the time series to obtain $\tilde{\Pi}_j$; the results are summarized in Figs.~\ref{fig:FT}-\ref{fig:FT_cascade}.

In Fig.~\ref{fig:FT} we show the Fourier transform of the conjugate momentum, $\tilde{\Pi}_j$ for $j=1,7,11$. The Fourier-transformed quantities show a clear excitation of the modes~\ref{modesD0} and \ref{modesN0}.
For a massless Dirichlet field (top panels of Fig.~\ref{fig:FT}), initially only the first modes are excited and the dominant frequency is $\omega R\sim 15$, whereas immediately before BH formation the frequency content seems to be approximately the same for all high-frequency bins satisfying $\omega R\leq 50$, i.e., all high frequencies are excited to roughly the same amplitude (cf. top right panel), thus suggesting that weak turbulence is at work in this case. On the other hand, different boundary conditions and the absence of a fully-resonant spectrum seem to change this behavior qualitatively, as shown in the central and bottom panels of Fig.~\ref{fig:FT} for massless Neumann and massive Dirichlet fields, respectively. In all cases, the peaks of the Fourier transform correspond to the eigenfrequencies~\eqref{modesD0} and \eqref{modesN0}, and in the case of massive fields the field is suppressed when $\omega<\mu$, as expected.

The transfer to higher frequencies is manifest in Fig.~\ref{fig:FT_cascade}, which shows the difference $\Delta\tilde{\Pi}_{j,j-1}\equiv \tilde{\Pi}_j-\tilde{\Pi}_{j-1}$ in the Fourier transform of $\Pi$ between two consecutive and equispaced time intervals. Focusing on the high-frequency region, $40\leq\omega R\leq100$, we consider the difference at three time intervals (from left to right) before the formation of an apparent horizon. In the massless Dirichlet case (top panels) the difference is always positive and the eigenmodes are clearly visible, signaling a clear \emph{direct cascade} effect where the frequency content of the field is shifted to higher and higher frequencies. On the other hand, the Neumann case (central panels) and the massive Dirichlet case (bottom panels) show a nontrivial pattern, where the eigenmodes are not noticeable. In some time intervals a direct cascade mechanism is at work, but the latter can be also followed by an \emph{inverse cascade} which shifts the field content to lower frequencies. Furthermore, in the case of a nonresonant spectrum the frequency transfer for a given time interval is also frequency dependent, as shown by the change of sign of $\Delta\tilde{\Pi}$ as a function of $\omega$ in some of the central and bottom panels of Fig.~\ref{fig:FT_cascade}.

A similar inverse cascade effect has been recently shown to halt collapse for a large family of initial data in AdS because it is competitive to weak turbulence~\cite{Balasubramanian:2014cja} (but see also~\cite{Bizon:2014bya}). However, our initial data are generic and do not necessarily belong to the class discussed in Ref.~\cite{Balasubramanian:2014cja}, which leads to dynamical, nondispersive solutions. In our case even when weak turbulence does not seem to be effective the end point of the simulations is the formation of an apparent horizon, suggesting that some other nonlinear mechanism (different from mode cascade triggered by a fully-resonant spectrum) might play a role in the gravitational collapses.

Our analysis is based on a Fourier transform of a time data series extracted at the origin. This approach seems to be more natural than a decomposition in Minkowski modes (cf. e.g. Ref.~\cite{Maliborski:2012gx}) to analyze the full nonlinear evolution and the approach to BH formation.
When using a decomposition in Minkowski modes, we recover results similar to the ones reported in Ref.~\cite{Maliborski:2012gx}.

\subsection{Relation with previous results in the literature}\label{sec:relation}
There appears to be some tension between our results and previous analysis identifying the commensurability of the spectrum of linear perturbations as a necessary condition for the nonlinear instability of AdS~\cite{Bizon:2011gg,Dias:2011ss,Dias:2012tq} and of other confining geometries~\cite{Maliborski:2014rma}. In a nutshell, the argument relies on the fact that BH formation is a nonlinear process which requires a breakdown of the perturbative analysis. Thus, BH formation for \emph{arbitrarily} small initial data requires a breakdown of the perturbative analysis even for arbitrarily small amplitudes. This can be accomplished if secular terms appear in the expansion, as when the spectrum is fully resonant~\cite{Bizon:2011gg,Dias:2012tq}.

On the other hand, although strong arguments can be made supporting that the perturbative series has a finite radius of convergence~\cite{Dias:2012tq}, it is not a priori guaranteed that the full nonlinear dynamics of the problem is captured by a perturbative analysis. In other words, some nonlinear interactions might not be described perturbatively. In fact, recent work even questions the validity of the standard perturbative expansions, which breaks down at late time~\cite{Balasubramanian:2014cja}. Thus, even if the perturbative analysis is well-behaved order by order, the nonlinear evolution might still lead to BH formation. This seems to be the case in the systems considered in this work, whose spectrum is not fully resonant and yet, gravitational collapse ensues for (arbitrarily?) small initial data.

Clearly, such tension would be resolved by assuming that a critical amplitude exists but is smaller than the minimum amplitude considered in our simulations. While this might always be the case -- even in a fully-resonant situation -- we note here that: (i) to the best of our knowledge, no argument can currently explain the behavior shown in Figs.~\ref{fig:coltime_Dirichlet} and~\ref{fig:coltime_DvsN}, namely why a \emph{dispersive} spectrum (either Dirichlet boundary in the massive case or Neumann boundary in the massless case) facilitates the collapse relative to the case of a fully-resonant spectrum with the same initial data; (ii) even if a critical amplitude turns out to exist, the question about the universality of the nonlinear instability might become an academic one: as shown in Figs.~\ref{fig:coltime_Dirichlet} and \ref{fig:coltime_DvsN}, gravitational collapse can occur for amplitudes which are orders of magnitude smaller than the critical amplitude for prompt collapse and it occurs after more than $10^5$ reflections, i.e. on extremely long time scales. Also this phenomenon does not entirely fit in the current understanding of the process.

To date the only simulations of a totally confining geometry (i.e. with no horizon and with no dissipation at infinity) and whose linearized spectrum is dispersive are those performed in flat spacetime~\cite{Maliborski:2012gx,Maliborski:2014rma} or in a portion of AdS~\cite{Buchel:2013uba}. While the results of Ref.~\cite{Maliborski:2012gx} are consistent with ours -- in the sense that Neumann boundary conditions were reported to yield qualitatively similar results to its Dirichlet counterpart -- Ref.~\cite{Maliborski:2014rma} recently reported some simulations which do not lead to gravitational collapse after $\sim4\times 10^3$ reflections. This fact has lead the authors of Ref.~\cite{Maliborski:2014rma} to conclude that, when the spectrum of linear perturbations is dispersive, there exists a critical amplitude below which the collapse is halted. Aside from the different choice of coordinates and formulation, another obvious difference between our simulations and that of Refs.~\cite{Maliborski:2012gx,Maliborski:2014rma} are the different initial data. Even in AdS, there exist some classes of initial data which are nonlinearly stable and do not lead to BH formation for arbitrarily small initial amplitude~\cite{Maliborski:2013jca,Buchel:2013uba} (see also Ref.~\cite{Balasubramanian:2014cja}, where a large family of such initial data has been constructed). Therefore, it is possible that our choice of initial data does not pass through any ``islands of stability'' when the amplitude is decreased, whereas the initial data used in Refs.~\cite{Maliborski:2012gx,Maliborski:2014rma} might do so. However, to test this hypothesis, we have studied two different sets of initial data, namely Eqs.~\eqref{eq:initPi_n} and \eqref{eq:initPi_a}, and we have considered different pulse parameters for each set. As previously discussed, we did not observe any qualitatively different feature depending on the initial data, even if the choice~\eqref{eq:initPi_n} is (only qualitatively) similar to the choice made in Refs.~\cite{Maliborski:2012gx,Maliborski:2014rma}. Nevertheless, these results are all pointing
towards faster collapse in the presence of dispersive spectrum, which is consistent with the results we report here and not favored by the current picture of weakly-turbulent collapse triggered by a fully-resonant spectrum.

Another possibility is that, as discussed above, even in our case there exists a critical amplitude which is smaller than the range considered here. Nonetheless, if this is the case, it would be interesting to understand why in our simulations the putative critical amplitude is so much smaller than in the case considered in Refs.~\cite{Maliborski:2012gx,Maliborski:2014rma}. On the other hand, it is also possible that no critical amplitude exists even in the case studied in Refs.~\cite{Maliborski:2012gx,Maliborski:2014rma} and that the evolutions shown in Ref.~\cite{Maliborski:2014rma} would eventually collapse on time scales which are longer than the one considered in that work. 

We stress here that it is essentially impossible to argue in favor of or against collapse with a few isolated simulations with different amplitudes: in the absence of a quantitative prediction to test, those simulations that do not collapse could always do so on longer time scales, whereas one could always argue that for those simulations that do bring to BH formation an even smaller amplitude would halt the collapse. 
In this view, we note that, by extrapolating a fit of the form $t_{\rm BH}\sim A^{-2}$ from the data shown in Ref.~\cite{Maliborski:2012gx}, one would predict that the collapse time for the amplitude $\epsilon=1/2$ considered in Ref.~\cite{Maliborski:2014rma} is roughly $t_{\rm BH}~\gtrsim 10^4 R$, which is much longer than the time interval shown in Fig.~1 of the same paper. This is particularly relevant in light of our Fig.~\ref{fig:curvature} which shows that at early times ($t\lesssim 10^3$) the growth of the curvature invariants is very mild or even absent.

In summary, it is possible that for very small amplitudes even simulations that seem to approach a stable bouncing configuration after thousands of reflections can in fact still grow unbound and collapse on even longer time scales\footnote{In this context, it is relevant to point out that the collapse time can be strongly delayed even in asymptotically-flat spacetime. For example, a scalar self-potential can confine low-frequency modes and lead to metastable ``oscillaton'' configurations that collapse only after several reflections off the potential barrier~\cite{Okawa:2013jba}.}.

\begin{figure}[ht]
 \psfig{file=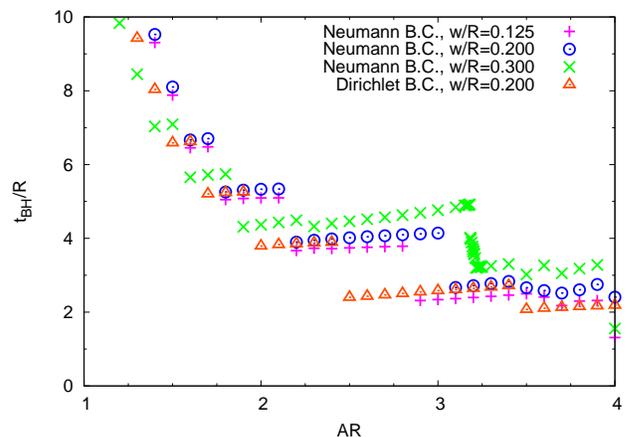,width=8.5cm}
 \caption[]{Example of the time of apparent horizon formation as a
 function of the amplitude in the range of moderately large amplitudes
 with initial profile~\eqref{eq:initPi_a}.
 The function displays an upward-stairway behavior for decreasing amplitudes which is connected to the power-law behavior shown in Fig.~\ref{fig:coltime_Dirichlet} at smaller amplitudes.
 \label{fig:upward}
 }
\end{figure}

Finally, the above picture is made even more complicated by the nontrivial behavior of the collapse time as a function of the amplitude. Figure~\ref{fig:upward} shows that, for moderately large amplitudes that are nonetheless smaller than Choptuik's critical value, the function $t_{\rm BH}(A)$ displays an upward-stairway behavior before drifting to the power-law growth shown in Fig.~\ref{fig:coltime_Dirichlet} at smaller amplitudes (see also Ref.~\cite{Buchel:2013uba} for a similar behavior in AdS). Thus, if one incidentally explores one of the plateau regions shown in Fig.~\ref{fig:upward}, one would be tempted to conclude that the collapse time saturates. This is another motivation to span a large range in amplitude as the one shown in Figs.~\ref{fig:coltime_Dirichlet} and \ref{fig:coltime_DvsN}.

%

\section{Do we understand the nonlinear instability of confined geometries?}
Even within the highly-simplifying assumption of spherical symmetry, the final state of the evolution of small perturbations in a confined geometry within full General Relativity is an extremely challenging problem. Due to finite time and computational resources, any approach based on numerical simulations is doomed to provide only partial answers or at most compelling evidence for/against the collapse.
In particular, performing single simulations for some (seemingly) small initial amplitude is likely not to provide any answer, as even smaller amplitudes could in principle contradict the result of the previous simulations. 

We stress here the importance of extracting the behavior of the collapse time as a function of the amplitude in a large domain, as shown in Figs.~\ref{fig:coltime_Dirichlet} and \ref{fig:coltime_DvsN} (cf. also Figs.~7 and 8 in Ref.~\cite{Buchel:2013uba}). Obviously --~in the absence of a solid analytical understanding of the process~-- even in this case an extrapolation to even smaller amplitudes is nevertheless required. In particular, it might be possible that a critical amplitude exists for some of the curves shown in Figs.~\ref{fig:coltime_Dirichlet} and \ref{fig:coltime_DvsN}, but that such amplitude is smaller than the range we considered. We note that this possibility is not ruled out even for the case of fully-resonant spectrum, for which only numerical simulations at finite amplitude are available (see, however, the recent perturbative approach developed in Ref.~\cite{Balasubramanian:2014cja}).

Our results support the idea that some families of initial data exist for which gravitational collapse for arbitrarily small initial amplitude occurs \emph{generically}, and not only for those systems that correspond to a fully-resonant spectrum of linear perturbations. However, for nonresonant configurations, a novel mechanism seems to exist that drives the system towards BH formation, and which is qualitatively different from weak turbulence driven by a fully-resonant spectrum. These claims are supported by four main findings:

\begin{enumerate}
\item For small amplitudes the collapse time is well approximated by $t_{\rm BH}\sim A^{-2}$ and we did not observe deviations from this power-law behavior\footnote{A power-law behavior as the one discussed above allows to \emph{estimate} the collapse time for small amplitudes, thus allowing future simulations to test this prediction and also helping to understand which initial data are expected to collapse within a given evolution time. 
For example, our results suggest that to disprove the power-law behavior shown in Fig.~\ref{fig:coltime_DvsN} using $A R\sim 10^{-4}$ (i.e. after decreasing by two orders of magnitude the smallest amplitude we considered), one should necessarily perform precise evolutions including more than $10^9$ reflections.} in any of our simulations.

\item Whether or not a critical amplitude exists for nonresonant configurations, we found that in the massless Neumann case and in the massive Dirichlet case the collapse time is always \emph{shorter} than in the massless Dirichlet case for the same initial data (cf. Figs.~\ref{fig:coltime_Dirichlet} and \ref{fig:coltime_DvsN}), a result which is at odds with the current understanding of the phenomenon in terms of weak turbulence driven by a resonant spectrum, because the two former cases are associated with a dispersive spectrum. 

\item While for fully-resonant configurations the collapse is driven by direct cascading of energy to higher frequencies, for nonresonant configurations we observed both direct-cascade effects and inverse-cascading of energy to lower frequencies (cf. Fig.~\ref{fig:FT_cascade}). The latter effect can quench turbulent instability for some configurations~\cite{Balasubramanian:2014cja} (but see~\cite{Bizon:2014bya}). However, our results leave open the possibility that even inverse cascade is not enough to prevent collapse at very late times for generic families of initial data. Furthermore, in the fully-resonant case the collapse is mode-driven, whereas no imprint of the linearized spectrum is observed for nonresonant configurations before BH formation.

\item Finally, in a large region of the initial-data parameter space the collapse time is well approximated by $t_{\rm BH}\sim \frac{R}{A^2w^2}$ (cf. Fig.~\ref{fig:coefficient}). This shows that, for a fixed small amplitude, initial pulses with narrow profiles take longer to collapse. This behavior is the opposite as what observed in a prompt collapse~\cite{Choptuik:1992jv}, and suggests that in a confined geometry the main responsible for the delayed collapse after several reflections is the nonlinear late-time dynamics, rather than the initial mode/energy content.

\end{enumerate}

Admittedly, our results raise more questions than answers, and suggest that the nonlinear instability of confined geometries is still an open problem. We hope our work will stimulate further studies on this fascinating and fundamental topic.

\begin{acknowledgments}
We are indebted to \'Oscar Dias, Luis Lehner and Jorge Santos for a critical reading of the manuscript and for many useful suggestions.
VC and PP acknowledge interesting discussions with the participants of the
workshop ``Holographic vistas on Gravity and Strings'', YITP-T-14-1, held at the Yukawa Institute for Theoretical Physics.
We acknowledge financial support provided under the European Union's FP7 ERC Starting Grant ``The dynamics of black holes:
testing the limits of Einstein's theory'' grant agreement no. DyBHo--256667.
This research was supported in part by Perimeter Institute for Theoretical Physics. 
Research at Perimeter Institute is supported by the Government of Canada through 
Industry Canada and by the Province of Ontario through the Ministry of Economic Development 
$\&$ Innovation.
This work was also supported by the NRHEP 295189 FP7-PEOPLE-2011-IRSES Grant, by FCT-Portugal through projects
CERN/FP/123593/2011 and IF/00293/2013, and by the Intra-European Marie Curie fellowships aStronGR-2011-298297 and AstroGRAphy-2013-623439.
Computations were performed on the ``Baltasar Sete-Sois'' cluster at IST.
\end{acknowledgments}

\appendix
\section{CARTOON method}\label{app:cartoon}
Evaluating numerically the Laplacian operator $\Lap$ in spherical symmetry might lead to large numerical errors near the origin due to the presence of $1/r$ terms.
However, as discussed in Ref.~\cite{Alcubierre:1999ab}, in spherical symmetry the flat Laplacian in Cartesian coordinates can easily be computed with sufficient accuracy
by using variables only on the x-axis. In terms of finite differences, the Laplacian operator reads
\begin{eqnarray}
 \Lap u &=& u'' +\frac{2u'}{r}
  = \partial_{x}^2u +\partial_{y}^2u +\partial_{z}^2u\nonumber\\
 &=& \frac{u(x+\Delta h, y, z)+u(x-\Delta h, y, z)-2u(x, y, z)}{\Delta h^2} \nonumber\\
 & & +\frac{u(x, y+\Delta h, z)+u(x, y-\Delta h, z)-2u(x, y, z)}{\Delta h^2} \nonumber\\
 & & +\frac{u(x, y, z+\Delta h)+u(x, y, z-\Delta h)-2u(x, y, z)}{\Delta h^2} \nonumber\\
 &=& \frac{u(x+\Delta h, y, z)+u(x-\Delta h, y, z)-2u(x, y, z)}{\Delta h^2} \nonumber\\
 & &+\frac{4}{\Delta h^2}\left[u(\sqrt{x^2+\Delta h^2}, y, z)-u(x,y,z)\right],\nonumber\\
 &=& \frac{u_{j+1}+u_{j-1}-2u_{j}}{\Delta h^2}
  +\frac{4}{\Delta h^2}\left[\tu_{j} -u_{j}\right]\,,
\end{eqnarray}
where $\tu_j$ is determined by the 2nd order Lagrange interpolation as $u(\sqrt{x^2+\Delta h^2}, y, z)$.

\section{Code tests}\label{app:convergence}

\begin{figure*}[ht]
 \begin{tabular}{cc}
  \psfig{file=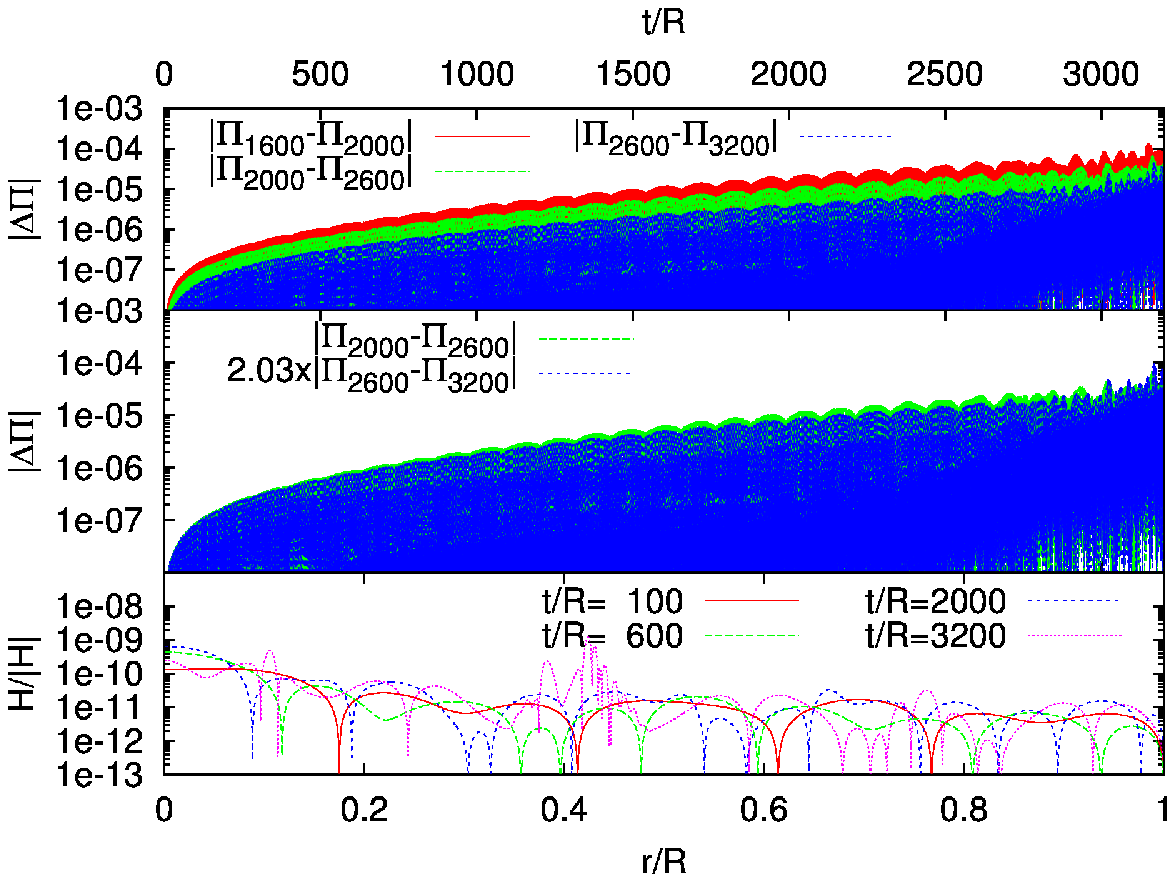,width=8.5cm}&
  \psfig{file=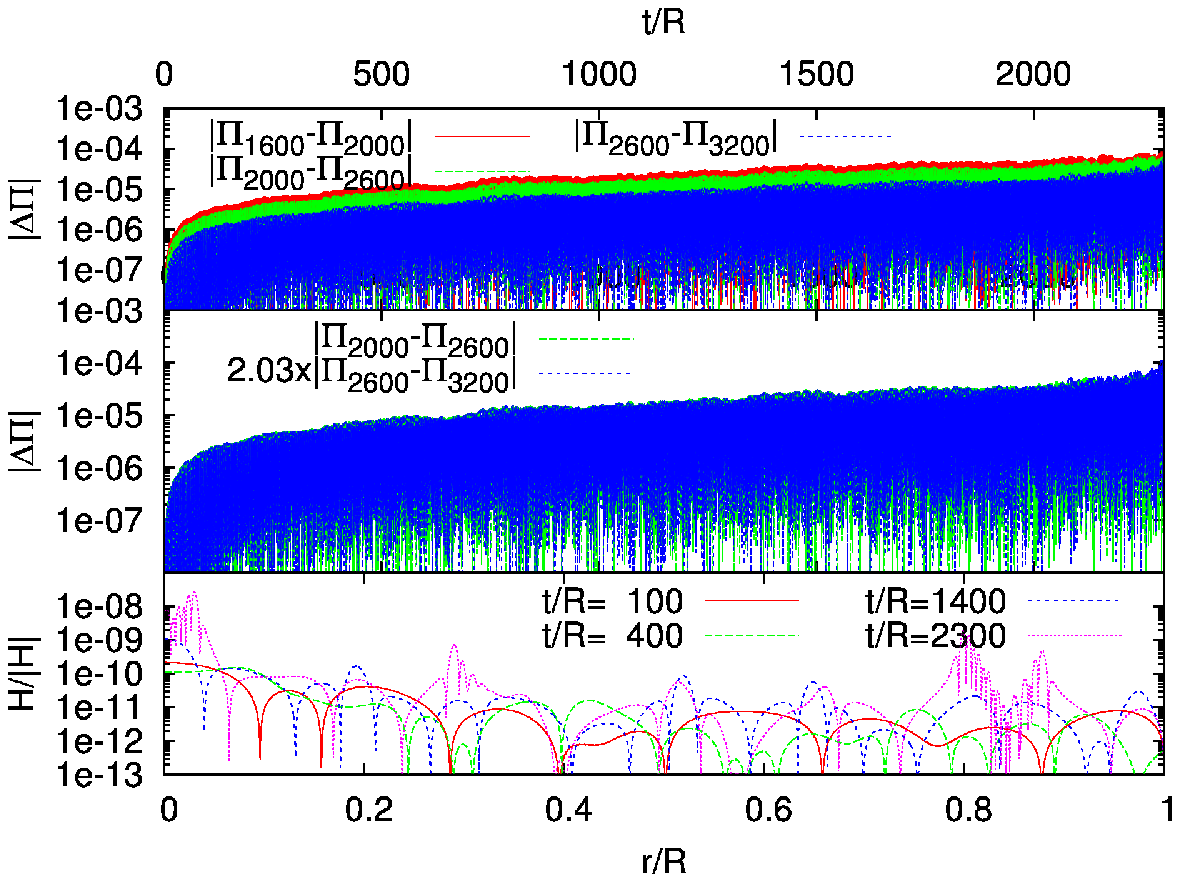,width=8.5cm}\\
  \psfig{file=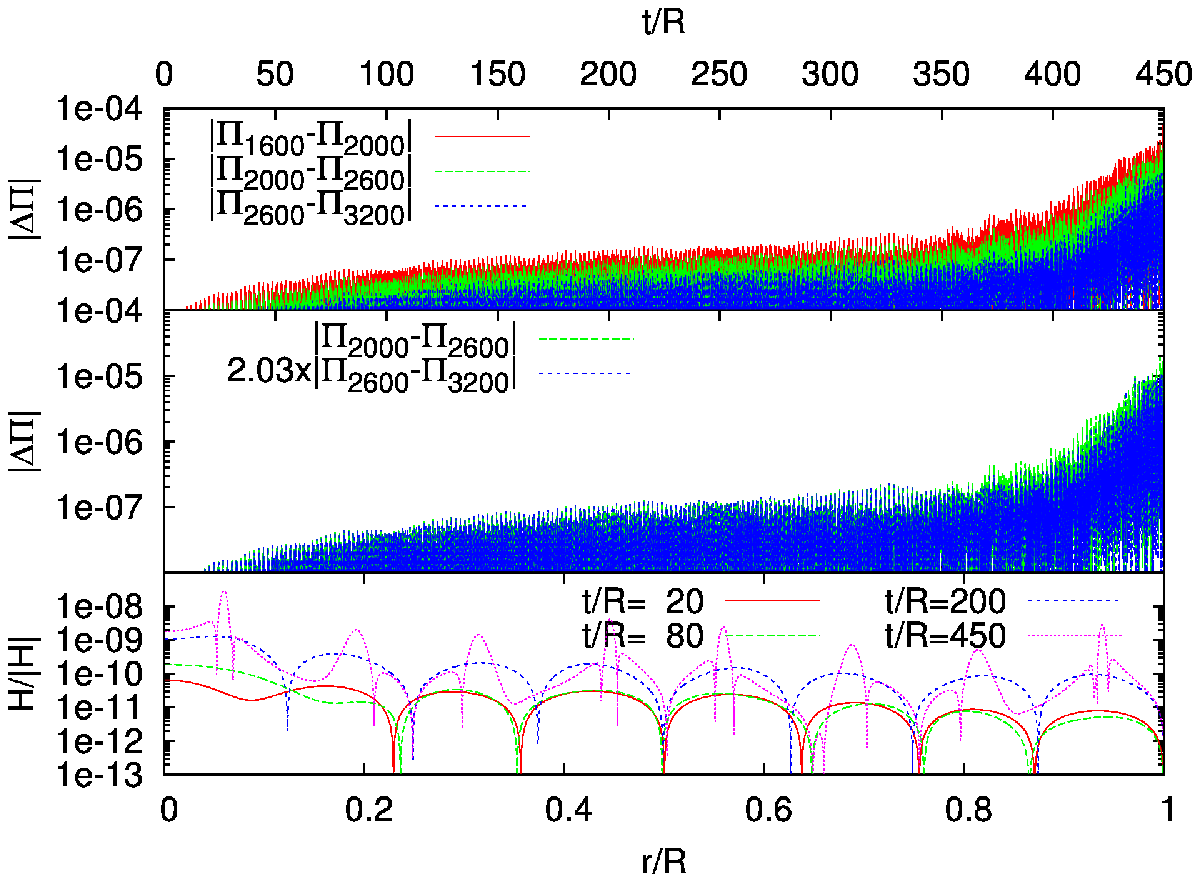,width=8.5cm}&
   \psfig{file=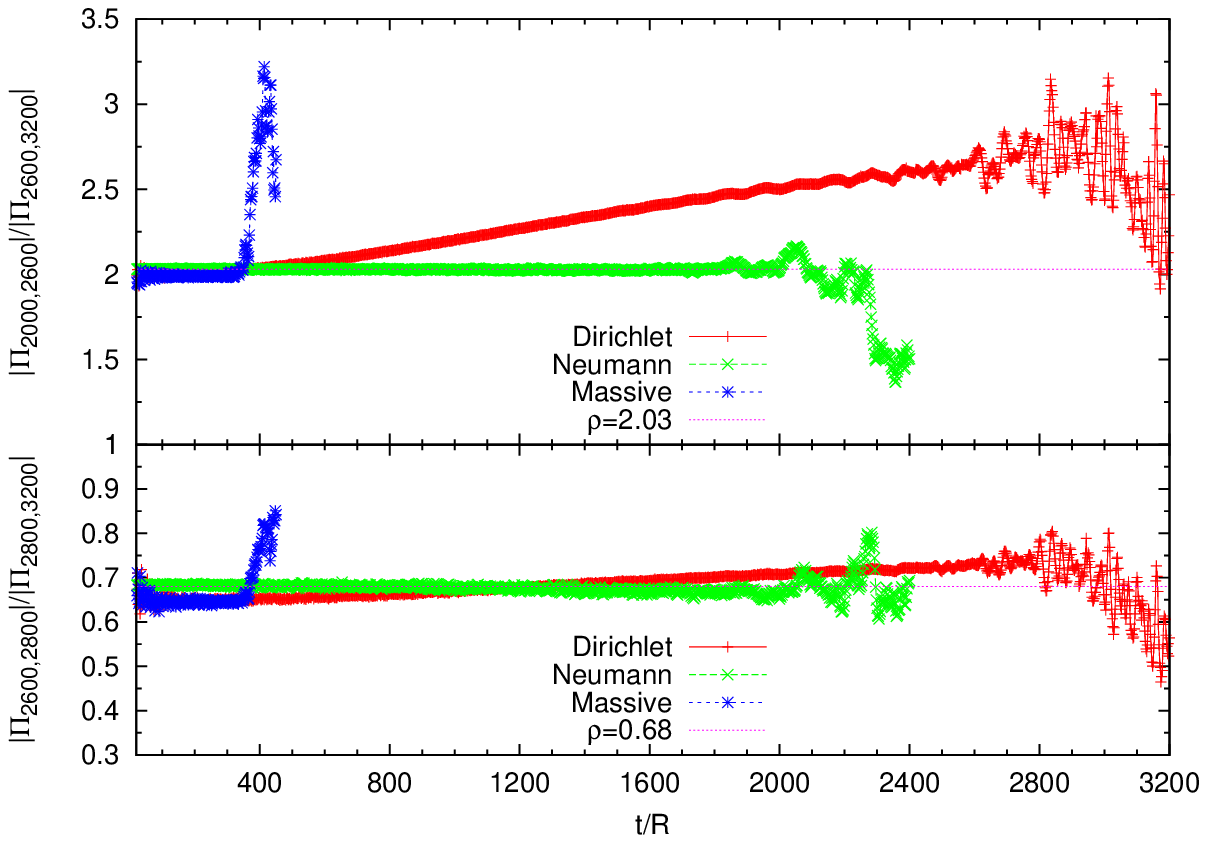,width=8.5cm}
 \end{tabular}
 \caption[]{Code tests with different boundary conditions and different mass terms. The top left plots, the top right plots and the bottom left plot correspond to Dirichlet boundary  conditions with $\mu R=0$, Neumann boundary conditions with $\mu R=0$ and Dirichlet boundary conditions with $\mu R=4$, respectively. Initial data corresponds to Eq.~\eqref{eq:initPi_n}. For each plot, the upper and middle panels show the convergence test for
 $\Pi$ at the center as a function of time, using different resolutions: $dr=R/1600, R/2000, R/2600$ and $R/3200$.
 Our results show second-order convergence as expected by the initial data solver.
 For each plot, the lower panel shows the Hamiltonian constraint as a function of coordinate $r$
 on different time slices with $dr=R/3200$ and normalized by the
 sum of the absolute value of each term in the Hamiltonian constraint.
 Bottom right plot: convergence factor computed through Eq.~\eqref{L2norm}.
 The upper panel shows the convergence factor $|\Pi_{2000,2600}|/|\Pi_{2600,3200}|$ using number of grid points $N=2000, 2600$ and
 $3200$, while lower panel shows $|\Pi_{2600,2800}|/|\Pi_{2800,3200}|$ using $N=2600, 2800$ and $3200$.
 All the plots refer to initial data set by Eq.~\eqref{eq:initPi_n} with $A R=0.04$ and $w=0.125R$.
 }
 \label{fig:conv_test}
\end{figure*}

Our results are convergent and consistent with the discretization order used.
In the first three plots of Fig.~\ref{fig:conv_test} we show the convergence analysis for the
initial data corresponding to the pulse~\eqref{eq:initPi_n} with different boundary conditions.
For each plot, the upper and middle panels display the comparison with different resolutions, showing second-order convergence.
The lower panels show the normalized Hamiltonian constraint as a function of the coordinate $r$ at different times.

Finally, in the bottom right plot of Fig.~\ref{fig:conv_test} we also show the actual convergence factor
extracted from the $L^2$-norm
\begin{equation}
 |\Pi_{N_1,N_2}(t_n,0)|\equiv\left[\sum_{t=nR}^{(n+2)R}(\Pi_{N_1}(t,0)-\Pi_{N_2}(t,0))^2\right]^{\frac{1}{2}}\,, \label{L2norm}
\end{equation}
for the quantity $\Pi$ at the center and at different time intervals over one period $T\sim 2R$ and where $N_i$ indicates the number of grid points, $N_i\equiv R/dr_i$ (cf. Fig.~2 in Ref.~\cite{Bizon:2014bya} for a similar convergence test). This definition allows the highly-oscillatory behavior of $\Pi$ at the center (as shown in the first three plots of Fig.~\ref{fig:conv_test}) to be removed.
Horizontal dashed lines in the bottom right plot of Fig.~\ref{fig:conv_test} denote the theoretically expected convergence factors computed by
\begin{equation}
 \rho = \frac{dr_{N_1}^{\gamma}-dr_{N_2}^{\gamma}}{dr_{N_2}^{\gamma}-dr_{N_3}^{\gamma}},
\end{equation}
where $\gamma$ denotes the expected order of convergence and therefore $\gamma=2$ in our case.
The upper and lower panels of this plot refer to two different resolutions, confirming that the expected convergence factor is achieved at sufficiently-high resolutions.
%

\bibliographystyle{h-physrev4}
\bibliography{collapse_box}

\end{document}